\documentclass[]{report}   

\usepackage{bm}
\usepackage[spanish,english]{babel}
\usepackage{amsfonts}
\usepackage{amssymb}
\usepackage{graphicx}
\usepackage[latin1]{inputenc}
\usepackage{hyperref}

\begin{document}

\title{Introduction\footnote{Chapter written for the book ``Aspects of Today's Cosmology", Edited by: Antonio Alfonso-Faus, InTech Publishing, Rijeka, Croatia, (2011), ISBN 978-953-307-626-3. The properly formatted version (26 pages) can be freely downloaded from the \href{http://www.intechopen.com/articles/show/title/introduction-to-modified-gravity-from-the-cosmic-speedup-problem-to-quantum-gravity-phenomenology}{publisher's website}.} to Modified Gravity: from the Cosmic Speedup Problem to Quantum Gravity Phenomenology }   
\author{Gonzalo J. Olmo\\ {\small Departamento de F\'{i}sica Te\'{o}rica \& IFIC }\\ {\small Universidad de Valencia - CSIC} \\ {\small Burjassot 46100, Valencia,} \\ {\small Spain}}         
\date{December, 2011}    

\maketitle

\begin{abstract}
 These notes represent a summary of the introductory part of a course on modified gravity delivered at several Spanish Universities (Granada, Valencia, and Valladolid), at the University of Wisconsin-Milwaukee (WI, USA), and at the Karl-Franzens Universit\"{a}t (Graz, Austria) during the period 2008-2011. We begin with a discussion of the classical Newtonian framework and how special relativity boosted the interest on new theories of gravity. Then we focus on Nordstr\"om's scalar theories of gravity and their influence on Einstein's theory of general relativity. We comment on the meaning of the Einstein equivalence principle and its implications for the construction of alternative theories of gravity. We present the cosmic speedup problem and how $f(R)$ theories can be constrained attending to their weak-field behavior. We conclude by showing that Palatini $f(R)$ and $f(R,Q)$ theories can be used to address different aspects of quantum gravity phenomenology and singularity problems. 
\end{abstract}

\addtocounter {chapter} {1}

\tableofcontents
\section{Introduction}

The reasons and motivations that lead to the consideration of alternatives
to General Relativity are manifold and have changed over the years. Some
theories are motivated by theoretical reasons while others are more phenomenological.
One can thus find theories aimed at unifying different interactions,
such as Kaluza-Klein theory (5-dimensional spacetime as a possible framework
to unify gravitation and electromagnetism) or the very famous string theory
(which should provide a unified explanation for {\it everything}, i.e., from particles
to interactions); others appeared as spin-offs of string theory and are now
seen as independent frameworks for testing some of its phenomenology, such
is the case of the string-inspired ``brane worlds'' (which confine the standard
model of elementary particles to a 4-dimensional  brane within a larger bulk accessible to
gravitational interactions); we also find modifications of GR needed to allow
for its perturbative renormalization, or modifications aimed at avoiding the
big bang singularity, effective actions related with non-perturbative quantization
schemes, etcetera.  All them are motivated by theoretical problems. \\
On the other hand, we find theories motivated by the need to find alternative explanations
for the current cosmological model and astrophysical observations,
which depict a Universe filled with some kind of {\it aether} or {\it dark energy}
representing the main part of the energy budget of the Universe, followed
by huge amounts of {\it unseen} matter which seems necessary to explain the
anomalous rotation curves of galaxies, gravitational lensing, and the formation
of structure via gravitational instability.

One of the goals of this chapter is to provide the reader with elementary
concepts and tools that will allow him/her better understand different
alternatives to GR recently considered in the literature in relation with the
cosmic speedup problem and the phenomenology of quantum gravity during the very
early universe. Since such theories are aimed at explaining certain
observational facts, they must be able to account for the new effects they
have been proposed for but also must be compatible with other observational
and experimental constraints coming from other scenarios. The process of
building and testing these theories is, in our opinion, a very productive theoretical
exercise, since it allows us to give some freedom to our imagination
but at the same time forces us to keep our feet on the ground. \\
Though there are no limits to imagination, experiments and observations should be used
as a guide to build and put limitations on sensible theories. In fact, a careful
theoretical interpretation of experiments can be an excellent guide to constrain
the family of viable theories. In this sense, we believe it is extremely
important to clearly understand the implications of the Einstein equivalence
principle (EEP). We hope these notes manage to convey the idea that
theorists should have a deep knowledge and clear understanding of the experiments
related with gravitation.

We believe that $f(R)$ theories of gravity are a nice toy model to study a
possible gravitational alternative to the {\it dark energy problem}. Their dynamics
is relatively simple and they can be put into correspondence with
scalar-tensor theories of gravity, which appear in many different contexts
in gravitational physics, from extended inflation and extended quintessence
models to Kaluza-Klein and string theory. On the other hand, $f(R)$ theories,
in the Palatini version, also seem to have some relation with non-perturbative
approaches to quantum gravity. Though such approaches have only been applied
with certain confidence in highly symmetric scenarios (isotropic and
anisotropic, homogeneous cosmologies) they indicate that the Big Bang singularity
can be avoided quite generally without the need for extra degrees
of freedom. Palatini $f(R)$ theories can also be designed to remove that
singularity and reproduce the dynamical equations derived from isotropic
models of Loop Quantum Cosmology.
Extended Lagrangians of the form $f(R,Q)$, being $Q$ the squared Ricci tensor, exhibit
even richer phenomenology than Palatini $f(R)$ models. These are very interesting
and promising aspects of these theories of gravity that are receiving increasing
attention in the recent literature and that will be treated in detail in
these lectures.

We begin with Newton's theory, the discovery of special relativity, and Nordstr\"{o}m's scalar theories as a way to motivate the idea of gravitation as a curved space phenomenon. Once the foundations of gravitation have been settled, we shift our attention to
the predictions of particular theories, paying special attention
to $f(R)$ theories and some extensions of that family of theories. We show how
the solar system dynamics can be used to reconstruct the form of the gravity Lagrangian and
how modified gravity can be useful in modeling certain aspects of quantum gravity phenomenology.

\section{From Newtonian physics to Einstein's gravity. }

In his Principia Mathematica (1687) Newton introduced the fundamental three laws of classical mechanics:
\begin{itemize}
\item  If no net force acts on a particle, then it is possible to select a set of reference frames (inertial frames), observed from which the particle moves without any change in velocity. This is the so called {\it Principle of Relativity} (PoR).
\item From an inertial frame, the net force on a particle of mass $m$ is $\vec{F}=m\vec{a}$.
\item Whenever a particle A exerts a force on another particle B, B simultaneously exerts a force on A with the same magnitude in the opposite direction.
\end{itemize}
Using Newton's laws one could explain all kinds of motion. When a nonzero force acts on a body, it accelerates at a rate that depends on its {\it inertial} mass $m_i$. A given force will thus lead to different accelerations depending on the inertial mass of the body. In his Principia, Newton also found an explanation to Kepler's empirical laws of planetary motion: between any two bodies separated by a distance $d$, there exists a force called {\it gravity} given by $F_g=G\frac{m_1m_2}{d^2}$. Here $G$ is a constant, and $m_1$, $m_2$ represent the {\it gravitational} masses of those bodies. When one studies experimentally Newton's theory of gravity quickly realizes that there is a deep relation between the inertial and the gravitational mass of a body. It turns out that the acceleration $a$ experienced by any two bodies on the surface of the Earth looks the same irrespective of the mass of those bodies. This suggests that inertial and gravitational mass have the same numerical values, $m_i=m_g$ (in general, they are proportional, being the proportionality constant the same for all bodies). This observation is known as {\it Newton's equivalence principle} or weak equivalence principle. \\
From Newton's laws it follows that Newtonian physics is based on the idea of {\it absolute space}, a background structure with respect to which accelerations can be effectively measured. However, the PoR implies that, unlike accelerations, absolute positions and velocities are not directly observable. This conclusion was challenged by some results published in $1865$ by J.C. Maxwell. In Maxwell's work, the equations of the electric and magnetic field were improved by the addition of a new term (Maxwell's displacement current). The new equations predicted the existence of electromagnetic waves. The explicit appearance in those equations of a speed $c$ was interpreted as the { existence of a privileged reference frame}, that of the {\it luminiferous aether}\footnote{The aether was supposed to have very special properties, such as a very high elasticity, and to exhibit no friction to the motion of bodies through it.}. According to this, it could  be possible to measure absolute velocities (at least with respect to the aether\footnote{The aether was assumed to be at rest because otherwise the light  from distant stars would suffer  distortions in their propagation due to local motions of this fluid.}). \\
This idea motivated the experiment carried out by Michelson and Morley in $1887$  to measure the relative velocity of the Earth in its orbit around the sun with respect to the aether\footnote{Note that the speed of sound is relative to the wind. Analogously, it was thought that the speed of light should be measured with respect to the aether. Due to the motion of the Earth, that speed should depend on the position of the Earth and the direction of the light ray. The interferometer was built on a rotating surface such that the full experiment could be rotated to observe periodic variations of the interference pattern.}. Despite the experimental limitations of the epoch, their experiment had enough precision to confirm that the speed of light is independent of the direction of the light ray and the position of the Earth in its orbit. \\
Motivated by this intriguing phenomenon, in $1892$ Lorentz proposed that moving bodies contract in the direction of motion according to a specific set of transformations. In $1905$ Einstein presented its celebrated theory of special relativity and derived the Lorentz transformations using the PoR and the observed constancy of the speed of light without assuming the presence of an aether. Therefore, though the principle of relative motion had been put into question by electromagnetism, it was salvaged by Einstein's reinterpretation\footnote{It is worth noting that Einstein's results did not rule out the aether, but they implied that its presence was irrelevant for the discussion of experiments.}. \\
As of that moment, it was understood that any good physical theory should be adapted to the new PoR. Fortunately, Minkowski $(1907)$ realized that Lorentz transformations could be nicely interpreted in a four dimensional space-time (he thus invented the notion of spacetime as opposed to the well-known spatial geometry of the time). In this manner, a Lorentz-invariant theory should be constructed using geometrical invariants such as scalars and four-vectors, which represents a geometrical formulation of the PoR.

\subsection{A relativistic theory of gravity: Nordstr\"{o}m's theory.}

The acceptance of the new PoR led to the development of relativistic theories of gravity in which the gravitational field was represented by different types of fields, such as scalars (in analogy with Newtonian mechanics) or vectors (in analogy with Maxwell's electrodynamics). A natural proposal\footnote{Another very natural proposal would be a relativistic theory of gravity inspired by Maxwell's electrodynamics, being $F_\mu\equiv m du_\mu/d\tau=k G_{\mu\nu}u^\nu$ with $G_{\mu\nu}=-G_{\nu\mu}$. Such a proposal immediately implies that $F_\mu u^\mu=0$ and is compatible with the constancy of $c^2$. } in this sense consists on replacing the Newtonian equations by the following relativistic versions \cite{Norton}
\begin{equation} \nabla^2\phi=4\pi G\rho \ \to \ \Box\phi=4\pi G\rho  \label{eq:gravity}
\end{equation}
\begin{equation} \frac{d\vec{v}}{dt}=-\vec{\nabla}\phi \ \ \to \ \frac{du_\mu}{d\tau}=-\partial_\mu\phi  \label{eq:matter}
\end{equation}
This proposal, however, is unsatisfactory. From the assumed constancy of the speed of light, $\eta_{\mu\nu}u^\mu u^\nu=-c^2$, one finds that $u_\mu \frac{du^\mu}{d\tau}=0$, which implies the unnatural restriction $u^\mu\partial_\mu \phi=\frac{d\phi}{d\tau}=0$, i.e., the gravitational field should be constant along any observer's world line. \\
To overcome this drawback, Nordstr\"{o}m proposed that the mass of a body in a gravitational field could vary with the gravitational potential \cite{Nord1} . Nordstr\"{o}m proposed a relativistic scalar theory of gravity in which the matter evolution equation (\ref{eq:matter}) was modified to make it compatible with the constancy of the speed of light
\begin{equation}
F_\mu\equiv \frac{d(m u_\mu)}{d\tau}=-m\partial_\mu \phi \ \leftrightarrow \ m\frac{du_\mu}{d\tau}+u_\mu \frac{d m}{d\tau}=-m\partial_\mu \phi .
\end{equation}
This equation implies that in a gravitational field $m$ changes as $m d\phi/d\tau=c^2dm/d\tau$, which leads to $m=m_0 e^{\phi/c^2}$ and avoids the undesired restriction $d\phi/dt=0$ of the theory presented before\footnote{Varying speed of light theories may also avoid the restriction $d\phi/d\tau=0$, but such theories  break the essence of special relativity by definition.}. The matter evolution equation can thus be written as
\begin{equation}
\frac{d u_\mu}{d\tau}=-\partial_\mu \phi -\frac{d \phi}{d\tau} u_\mu \ .
\end{equation}
It is worth noting that this equation satisfies Newton's equivalence principle in the sense that the gravitational mass of a body is identified with its rest mass. Free fall, therefore, turns out to be independent of the rest mass of the body. However, Einstein's special theory of relativity had shown a deep relation between mass and energy that should be carefully addressed in the construction of any relativistic theory of gravity. The equation $E=mc^2$, where $m=\gamma m_0$ and $\gamma=1/\sqrt{1-\vec{v}^2/c^2}$, states that kinetic energy increases the effective mass of a body, its inertia. Therefore, if inertial mass is the source of the gravitational field, a moving body could generate a stronger gravitational field than the same body at rest. By extension of this reasoning, one can conclude that bodies with different internal energies could fall differently in an external gravitational field. Einstein found this point disturbing and used it to criticize Nordstr\"{o}m's theory. In addition, in this theory the gravitational potential $\phi$ of point particles goes to $-\infty$ at the location of the  particle, thus implying that point particles are massless and, therefore, cannot exist. One is thus led to consider extended (or continuous) objects, which possess other types of inertia in the form of stresses that cannot be reduced to a mass. The source of the gravitational field, the right hand side of (\ref{eq:gravity}), should thus take into account also such stresses. \\
To overcome those problems and others concerning energy conservation pointed out by Einstein, Nordstr\"{o}m  proposed a second theory \cite{Nord2}
\begin{equation}  \Box\phi=g(\phi)\nu  \label{eq:gravity2}
\end{equation}
\begin{equation} \mathcal{F}_\mu =-g(\phi)\nu \partial_\mu\phi  \label{eq:matter2} \ .
\end{equation}
where $\mathcal{F}$ represents the force per unit volume and $g(\phi)\nu$ is a density that represents the source of the gravitational field. To determine the functional form of $g(\phi)$ and find a natural correspondence between $\nu$ and the matter sources, Nordstr\"{o}m proceeded as follows. Firstly, he defined the {\it gravitational mass} of a system using the right hand side of (\ref{eq:gravity2}) and (\ref{eq:matter2}) as
\begin{equation} M_g=\int d^3x g(\phi)\nu  \ . \label{eq:Mg}
\end{equation}
Then he assumed that the {\it inertial mass} of the system should be a Lorentz scalar made out of all the energy sources, which include the rest mass and stresses associated to the matter, the gravitational field, and the electromagnetic field. He thus proposed the following expression
\begin{equation} m_i=-\frac{1}{c^2}\int d^3x[T_m+G_\phi+F_{em}] \ , \label{eq:mi}
\end{equation}
where the trace of the stress-energy tensor of the matter is represented by $T_m$, the trace of the electromagnetic field by $F_{em}$ (which vanishes), and that of the gravitational field by $G_m$, being $G_{\mu\nu}=(2/\kappa^2)[\partial_\mu\phi\partial_\nu\phi-(1/2)\eta_{\mu\nu}(\partial_\lambda\phi\partial^\lambda\phi)]$ the stress-energy tensor of the (scalar) gravitational field. \\
To force the equivalence between inertial and gravitational mass in a system of particles immersed in an external gravitational field with potential $\phi_a$, Nordstr\"{o}m imposed that for such a system the following relation should hold
\begin{equation} M_g=g(\phi_a)m_i \ . \label{eq:Mg=gmi}
\end{equation}
Then he considered a stationary system on that gravitational field and showed that the contribution of the local gravitational field to the total inertia of the system was given by
\begin{equation} -\frac{1}{c^2}\int d^3x G_\phi=-\frac{1}{c^2}\int d^3x (\phi-\phi_a)g(\phi)\nu \ .
\end{equation}
Combining this expression with (\ref{eq:Mg=gmi}) and (\ref{eq:mi}) one finds that
\begin{equation} \int d^3x\left[T_m+g(\phi)\nu\left(\phi-\phi_a+\frac{c^2}{g(\phi_a)}\right)\right] =0 \ .
\end{equation}
Demanding proportionality between $T_m$ and $\nu$, one finds that $g(\phi)=C/(A+\phi)$. A natural gauge corresponds to $g(\phi)=-4\pi G/\phi$ because it allows to recover the Newtonian result $E_0=m c^2=M_g\phi_a$ that implies that the energy of a system with gravitational mass $M_g$ in a field with potential $\phi_a$ is exactly $M_g\phi_a$. Therefore, from Nordstr\"{o}m's second theory it follows that the inertial mass of a stationary system varies in proportion to the external potential whereas $M_g$ remains constant, i.e., $m/\phi= constant$. \\
With the above results one finds that (\ref{eq:gravity2}) and (\ref{eq:matter2}) turn into (from now on $\kappa^2\equiv 8\pi G$)
\begin{equation}  \phi\Box\phi=-\frac{\kappa^2}{2}T_m  \label{eq:gravity3}
\end{equation}
\begin{equation} \frac{du_\mu}{d\tau} =-\partial_\mu\ln\phi-u_\mu\frac{d}{d\tau}\ln \phi  \label{eq:matter3} \ .
\end{equation}
Using these equations it is straightforward to verify that the total energy-momentum of the system is conserved, i.e.,  $\partial^\mu\left(T_{\mu\nu}^\phi+T_{\mu\nu}^m\right)=0$, where one must take $T_{\mu\nu}^m=\rho\phi u_\mu u_\nu$ for pressureless matter because, as shown above, the inertial rest mass density of a system grows linearly with $\phi$. \\
Nordstr\"{o}m's second theory, therefore, represents a satisfactory example of relativistic theory of gravity in Minkowski space that satisfies the equivalence between inertial and gravitational mass and in which energy and momentum are conserved. Unfortunately, it does not predict any bending of light and also fails in other predictions that were important at the beginning of the twentieth century such as the perihelion shift of Mercury. Nonetheless, it admits a geometric interpretation that greatly simplifies its structure and puts forward the direction in which  Einstein's work was progressing.\\
Considering a line element of the form $ds^2=\phi^2(-dt^2+d\vec{x}^2)$, Einstein and Fokker showed that the matter evolution equation (\ref{eq:matter3}) could be obtained by extremizing the path followed by a free particle in that geometry, i.e., by computing the variation $\delta\left(-mc^2\int ds\right)=0$ \cite{Ein-Fok} . This variation yields the geodesic equation\footnote{To obtain (\ref{eq:matter3}) from the geodesic equation one should note that $d\tilde{\tau}=\phi d\tau$, $\tilde{u}_\mu=\phi u_\mu$, and that the indices in (\ref{eq:matter3}) are raised and lowered with $\eta_{\mu\nu}$.}
\begin{equation}\label{eq:geodesics} \frac{d\tilde{u}^\mu}{d\tilde{\tau}}+\Gamma^\mu_{\alpha\beta}\tilde{u}^\alpha\tilde{u}^\beta=0 \ ,
\end{equation}
where $\Gamma^\mu_{\alpha\beta}=\partial_\alpha\phi \delta^\mu_\beta+\partial_\beta\phi \delta^\mu_\alpha-\eta^{\mu\rho}\partial_\rho\phi \eta_{\alpha\beta}$. The gravitational field equation also takes a very interesting form
\begin{equation} R=3\kappa^2 \tilde{T}_m \ ,
\end{equation}
where $R=-(6/\phi^3)\eta^{\alpha\beta} \partial_\alpha\partial_\beta \phi$ and $\tilde{T}_m=T_m/\phi^4$ due to the conformal transformation that relates the background metric with the Minkowski metric.
These last results represent generally covariant equations that establish a non-trivial relation between gravitation and geometry. Though this theory was eventually ruled out by observations, its potential impact on the eventual formal and conceptual formulation of Einstein's general theory of relativity must have been important.

\subsection{To general relativity via general covariance}

The Principle of Relativity together with Newton's ideas about the equivalence between inertial and gravitational mass led Einstein to develop what has come to be called the Einstein equivalence principle (EEP), which will be introduced later in detail. Einstein wanted to extend the principle of relativity not only to inertial observers (special relativity) but to all kinds of motion (hence the term general relativity). This motivated the search for generally covariant equations \footnote{The idea of general covariance is nowadays naturally seen as a basic mathematical requirement in any theory based on the use of differential manifolds. In this sense, though general covariance forces the use of tensor calculus, it should be noted that it does not necessarily imply curved space-time. Note also that it is the connection, not the metric, the most important object in the construction of tensors.}. \\
Though it is not difficult to realize that one can construct a fully covariant theory in Minkowski space, the consideration of arbitrary accelerated frames leads to the appearance of inertial or ficticious forces whose nature is difficult to interpret. This is due to the fact that Minkowski spacetime, like Newtonian space, is an absolute space. The possibility of writing the laws of physics in a coordinate (cartesian, polar,\ldots ) and frame (inertial, accelerated,\ldots) invariant way, helped Einstein to realize that a local, homogeneous gravitational field is indistinguishable from a constant acceleration. This allowed him to introduce the concept of local inertial frame (LIF) and find a correspondence between gravitation and geometry, which led to a deep conceptual change: {\bf there exists no absolute space}. This follows from the fact that, unlike other well-known forces, the local effects of gravity can always be eliminated by a suitable choice of coordinates (Einstein's elevator). \\
The forces of Newtonian mechanics, which were thought to be measured with respect to {\it absolute space}, were in fact being measured in an accelerated frame (static with respect to the Earth), which led to the appearance of the observed gravitational acceleration. According to Einstein, accelerations produced by interactions such as the electromagnetic field should be measured in LIFs. This means that they should be measured not with respect to absolute space but with respect to the local gravitational field (which defines LIFs). In other words, Einstein identified the Newtonian absolute space with the local gravitational field. Physical accelerations should, therefore, be measured in local inertial frames, where Minkowskian physics should be recovered. Gravitation, according to Einstein, was intrinsically different from the rest of interactions. It was a geometrical phenomenon.

The geometrical interpretation of gravitation implied that it should be described by a tensor field, the metric $g_{\mu\nu}$, which boils down to the Minkowski metric locally in appropriate coordinate systems (LIFs) or globally when gravitation is absent. This view made it natural to interpret the effects of a gravitational field on particles as geodesic motion. In the absence of non-gravitational interactions, particles should follow geodesics of the background metric, which are formally described by eq.(\ref{eq:geodesics}) but with $\Gamma^\mu_{\alpha\beta}$, the so-called Levi-Civita connection, defined in terms of a symmetric metric tensor $g_{\mu\nu}$ as
\begin{equation} \Gamma^\mu_{\alpha\beta}=\frac{g^{\mu\rho}}{2}\left[\partial_\alpha g_{\rho\beta}+\partial_\beta g_{\rho\alpha}-\partial_\rho g_{\alpha\beta}\right] \ . \label{eq:LC}
\end{equation}
To determine the dynamics of the metric tensor one needs at least ten independent equations, as many as independent components there are in $g_{\mu\nu}$. Since the source of the gravitational field must be related with the stress-energy tensor of matter and the dynamics of classical mechanics is generally governed by second-order equations, Einstein proposed the following set of tensorial equations
\begin{equation}
R_{\mu\nu}-\frac{1}{2}g_{\mu\nu}R=\kappa^2 T_{\mu\nu} \ , \label{eq:GR}
\end{equation}
where $R_{\mu\nu}\equiv{R^\rho}_{\mu\rho\nu}$ is the so-called Ricci tensor, $R=g^{\mu\nu}R_{\mu\nu}$ is the Ricci scalar, and ${R^\alpha}_{\beta\mu\nu}=\partial_\mu\Gamma_{\nu\beta}^\alpha-\partial_\nu\Gamma_{\mu\beta}^\alpha+\Gamma_{\mu\lambda}^\alpha\Gamma_{\nu\beta}^\lambda-\Gamma_{\nu\lambda}^\alpha\Gamma_{\mu\beta}^\lambda$ represents the components of the Riemann tensor, the field strength of the connection $\Gamma^\alpha_{\mu\beta}$, which here is defined as in (\ref{eq:LC}). \\
Eq. (\ref{eq:GR}) represents a system of non-linear, second-order partial differential equations for the ten independent components of the metric tensor. The conservation of energy and momentum is guaranteed independently for the left and the right hand sides of (\ref{eq:GR}). The contraction\footnote{The differential operator $\nabla_\mu$ represents a covariant derivative, which is the natural extension of the usual flat space derivative $\partial_\mu$ to spaces with non-trivial parallel transport.} $\nabla^\mu (R_{\mu\nu}-\frac{1}{2}g_{\mu\nu}R)=0$ follows from a geometrical identity, whereas  $\nabla^\mu T_{\mu\nu}=0$ follows if the Minkowski equations of motion for the matter fields are satisfied locally. The non-linearity of the equations manifests the fact that the energy stored in the gravitational field can source the gravitational field itself in a non-trivial way. Unlike Nordstr\"{o}m's second theory, this set of tensorial equations imply that the gravitational field is sourced by the full stress-energy tensor, not just by its trace. This implies that electromagnetic fields, like any other matter sources, generate a non-zero Ricci tensor and, therefore, gravitate.

Einstein's theory was rapidly accepted despite its poor experimental verification. In fact, we had to wait until the 1960's to have the perihelion shift of Mercury and the deflection of light by the sun measured to within an accuracy of $\sim 1\%$ and $\sim 50\%$, respectively. In 1959 Pound and Rebka were able to measure the gravitational redshift for the first time. Additionally, though Hubble's discoveries on the recession of distant galaxies had boosted Einstein's popularity, those observations were a mere qualitative verification of the effect and only recently has it been possible to contrast theory and observations with some confidence in the cosmological setting. It is therefore not surprising that between 1905 and 1960, there appeared at least 25 alternative relativistic theories of gravitation, where spacetime was flat and gravitation was a Lorentz-invariant field on that background. Though many researchers defended Einstein's idea of curved spacetime, others like Birkhoff did not \cite{Birkhoff}:

\begin{center}
\begin{minipage}{0.9\textwidth}
\emph{\small The initial attempts to incorporate gravitational phenomena in flat space-time were not satisfactory. Einstein turned to the curved spacetime suggested by his principle of equivalence, and so constructed his general
theory of relativity. The initial predictions, based on this celebrated
theory of gravitation, were brilliantly confirmed. However, the theory has
not led to any further applications and, because of its complicated mathematical
character, seems to be essentially unworkable. {Thus curved spacetime
has come to be regarded by many as an auxiliary construct (Larmor)
rather than as a physical reality}.}
\end{minipage}
\end{center}

Such strong claims suggest that it was necessary a careful analysis of the foundations of Einstein's theory: is spacetime really curved or is gravitation a tensor-like interaction in a flat background? The next section is devoted to clarify these points and others that will help establish the foundations of gravitation theory.

\subsection{The Einstein equivalence principle}

The experimental facts that support the foundations of gravitation should never be underestimated since they provide a valuable guide in the construction of viable theories and in constraining the realm of speculation. In this sense, the experimental efforts carried out by Robert Dicke in the 1960's \cite{Dicke1964} resulted in what has come to be called the Einstein equivalence principle (EEP) and constitute a fundamental pillar for gravitation theory. We will briefly review next the experimental evidence supporting it, and the way it enters in the construction of gravitation theories \cite{Will93}. The EEP states that \cite{Will05}
\begin{itemize}
\item Inertial and gravitational masses coincide (weak equivalence principle).
\item The outcome of any non-gravitational experiment is independent of the
velocity of the freely-falling reference frame in which it is
performed (Local Lorentz Invariance).
\item The outcome of any local non-gravitational experiment is independent of where and
when in the universe it is performed (Local Position Invariance).
\end{itemize}

Let us briefly discuss the experimental evidence supporting the EEP.

\subsubsection{Weak equivalence principle}

A direct test of WEP is the comparison of the acceleration of two laboratory-sized bodies of different
composition in an external gravitational field. If the principle were violated, then the accelerations
of different bodies would differ. In Dicke's torsion balance experiment, for instance, the gravitational acceleration toward the sun of small gold and aluminum weights were compared and found to be equal with an accuracy of about a part in $10^{11}$. One should note that gold and aluminum atoms have very different properties, which is important for testing how gravitation couples to different particles and interactions. For instance, the electrons in aluminum are non-relativistic whereas  the k-shell electrons of gold have a $15\%$ increase in their mass as a result of their relativistic velocities. The electromagnetic negative contribution to the binding energy of the nucleus varies as $Z^2$ and represents $0.5 \%$ of the total mass of a gold atom, whereas it is negligible in Al. Additionally, the virtual pair field, pion field, etcetera,  around the gold nucleus would be expected to represent a far bigger contribution to the total energy than in aluminum. This makes it clear that a gold sphere possesses additional inertial contributions due to the electromagnetic, weak, and strong interactions that are not present (or are negligible) in the aluminum sphere. If any of those sources of inertia did not contribute by the same amount to the gravitational mass of the system, then gold and aluminum would fall with different accelerations. \\
The precision of Dicke's experiment was such that from it one can conclude, for instance, that positrons and other antiparticles fall down, not up  \cite{Dicke1964}. This is so because if the positrons in the pair field of the gold atom were to tend to fall up, not down, there would be an anomalous weight of the atom substantially greater for large atomic number than small.

\subsubsection{Tests of local Lorentz invariance}

The existence of a preferred reference frame breaking the local isotropy of space would imply a dependence of the speed of light on the direction of propagation. This would cause shifts in the energy levels of atoms and nuclei that depend on the orientation of the quantization axis of the state relative to our universal velocity vector, and on the quantum numbers of the state. This idea was tested by Hughes (1960) and Drever (1961), who examined the $J=3/2$ ground state of the $^7Li$ nucleus in an external magnetic field. If the Michelson-Morley experiment had found $\delta\equiv c^{-2}-1 \approx 10^{-3}$, the Hughes-Drever experiment set the limit to $\delta\approx 10^{-15}$. More recent experiments using laser-cooled trapped atoms and ions have reached $\delta\approx 10^{-17}$. \\
Currently, new ideas coming from quantum gravity (with a minimal length scale), braneworld scenarios, and models of  string theory have motivated new ways to test Lorentz invariance by considering Lorentz-violating parameters in extensions of the standard model and also some astrophysical tests. So far, however, no compelling evidence for a violation of Lorentz invariance has been found.

\subsubsection{Tests of local position invariance}

Local position invariance can be tested by gravitational redshift experiments, which test the existence of spatial dependence on the outcome of local non-gravitational experiments, and by measurements of the fundamental non-gravitational constants that test for temporal dependence. Gravitational redshift experiments usually measure the frequency shift $Z=\Delta \nu/\nu=-\Delta \lambda/\lambda$ between two identical frequency standards (clocks) placed at rest at different heights in a static gravitational field. If the frequency of a given type of atomic clock is the same when measured in a local, momentarily comoving freely falling frame (Lorentz frame), independent of the location or velocity of that frame, then the comparison of frequencies of two clocks at rest at different locations boils
down to a comparison of the velocities of two local Lorentz frames, one at rest with respect to one
clock at the moment of emission of its signal, the other at rest with respect to the other clock at
the moment of reception of the signal. The frequency shift is then a consequence of the first-order
Doppler shift between the frames. The result is a shift $Z=\frac{\Delta U}{c^2}$, where U is the difference in the Newtonian gravitational potential between the receiver and the
emitter. If the frequency of the clocks had some dependence on their position, the shift could be written as $Z=(1+\alpha)\frac{\Delta U}{c^2}$. Comparison of a hydrogen-maser clock flown on a rocket to an altitude of about $10.000$ km with a similar clock on the ground yielded a limit $\alpha< 2\times 10^{-4}$. \\
Another important aspect of local position invariance is that if it is satisfied then the fundamental
constants of non-gravitational physics should be constants in time. Though these tests are subject to many uncertainties and experimental limitations, there is no strong evidence for a possible spatial or temporal dependence of the fundamental constants.

\subsection{Metric theories of gravity}

The EEP is not just a verification that gravitation can be associated with a metric tensor which locally can be turned into the Minkowskian metric by a suitable choice of coordinates. If it is valid, then gravitation must be a curved space-time phenomenon, i.e., the effects of gravity must be equivalent to the effects of living in a curved space-time. For this reason, the only theories of gravity that have a hope of being viable are those that satisfy the following postulates (see \cite{Will93} and \cite{Will05}):
\begin{enumerate}
\item Spacetime is endowed with a symmetric metric.
\item The trajectories of freely-falling bodies are geodesics of that
metric.
\item In local freely-falling reference frames, the
non-gravitational laws of physics are those written in the
language of special relativity.
\end{enumerate}
Theories satisfying these postulates are known as {\it metric theories of gravity}, and their action can be written generically as
\begin{equation}\label{eq:MTG}
S_{MT}= S_{G}[{g}_{\mu \nu },\phi,A_{\mu },B_{\mu \nu
},\ldots] +S_m[{g}_{\mu \nu },\psi_m] \ ,
\end{equation}
where $S_m[{g}_{\mu \nu },\psi_m]$ represents the matter action, $\psi_m$ denotes the matter and non-gravitational fields, and $S_G$ is the gravitational action, which besides the metric $g_{\mu\nu}$ may depend on other gravitational fields (scalars, vectors, and tensors of different ranks). This form of the action guarantees that the non-gravitational fields of the standard model of elementary particles couple to gravitation only through the metric, which should allow to recover locally the non-gravitational physics of Minkowski space. The construction of $S_m[{g}_{\mu \nu },\psi_m]$ can thus be carried out by just taking its Minkowski space form and going over to curved space-time using the methods of differential geometry. It should be noted that the EEP does neither point towards GR as the preferred theory of gravity nor provides any constraint or hint on the functional form of the gravitational part of the action. The functional $S_G$ must provide dynamical equations for the metric (and the other gravitational fields, if there are any) but its form must be obtained by theoretical reasoning and/or by experimental exploration.

It is worth noting that if $S_{G}$ contains other long-range fields besides the metric, then {\bf gravitational} experiments in a local, freely falling frame may depend on the location and velocity of the frame relative to the external environment. This is so because, unlike the metric, the boundary conditions induced by those fields cannot be trivialized by a suitable choice of coordinates. Of course, local  {\bf non-gravitational} experiments are unaffected since the gravitational fields they generate are assumed to be negligible, and since those experiments couple only to the metric, whose form can always be made locally Minkowskian at a given spacetime event. \\
Before concluding this section, it should be noted that string theories predict the existence of new kinds of fields with couplings to fermions and the interactions of the standard model in a way that breaks the simplicity of metric theories of gravity, i.e., they do not allow for a clean splitting of the action into a matter sector plus a gravitational sector. Such theories, therefore, must be regarded as non-metric. Improved tests of the EEP could be used to test the presence and/or intensity of such couplings, which are expected to represent short range interactions. These tests can be seen as a branch of high-energy physics not based on particle accelerators.

\subsubsection{Two examples of metric theories: General relativity and Brans-Dicke theory.}

The field equations of Einstein's theory of general relativity (GR) can be derived from the following action
\begin{equation}\label{eq:action-GR}
S[{g}_{\mu \nu},\psi_m]=\frac{1}{16\pi G}\int d^4x\sqrt{-g} R(g)+S_m[g_{\mu\nu},\psi_m]
\end{equation}
where $R$ is the Ricci scalar defined below eq.(\ref{eq:GR}). Variation of this action with respect to the metric leads to Einstein's field equations\footnote{Recall that $\delta\sqrt{-g}=-\frac{1}{2}\sqrt{-g}g_{\mu\nu}\delta g^{\mu\nu}$ and that $\delta R_{\mu\nu}=-\nabla_\mu \delta\Gamma^\lambda_{\lambda \nu}+\nabla_\lambda \delta\Gamma^\lambda_{\mu \nu}$.}
\begin{equation}\label{Gmn}
R_{\mu\nu}-\frac{1}{2}g_{\mu\nu}R=8\pi G T_{\mu\nu}
\end{equation}
 In Einstein's theory, gravity is mediated by a rank-2 tensor field, the metric, and curvature is generated by the matter sources. Brans-Dicke theory  introduces, besides the metric, a new gravitational field, which is a scalar. This scalar field is coupled to the curvature as follows
\begin{equation} \label{eq:ST}
S[{g}_{\mu \nu},\phi,\psi_m]=\frac{1}{16\pi }\int d^4
x\sqrt{-{g}}\left[\phi {R}({g})-\frac{\omega}{\phi}(\partial_\mu \phi\partial^\mu\phi)-V(\phi)
\right]+S_m[{g}_{\mu \nu},\psi_m]
\end{equation}
In the original Brans-Dicke theory, the potential was set to zero, $V(\phi)=0$, so the theory had only one free parameter, the constant $\omega$ in front of the kinetic energy term, which had to be determined experimentally. Note that the Brans-Dicke scalar has the same dimensions as the inverse of Newton's constant and, therefore, can be seen as related to it. In Brans-Dicke theory, one can thus say that Newton's constant is no longer constant but is, in fact, a dynamical field. The field equations for the metric are
\begin{equation}\label{eq:Gab-ST}
R_{\mu \nu }(g)-\frac{1}{2}g_{\mu \nu }R(g)=
\frac{8\pi}{\phi}T_{\mu\nu}-\frac{1}{2\phi}g_{\mu\nu}V(\phi)+\frac{1}{\phi}\left[\nabla_\mu\nabla_\nu\phi-g_{\mu\nu}\Box \phi\right]+\frac{\omega}{\phi^2}\left[\partial_\mu\phi\partial_\nu\phi-\frac{1}{2}g_{\mu\nu}(\partial
 \phi)^2\right]
\end{equation}
The equation that governs the scalar field is
\begin{equation}\label{eq:phi-ST}
(3+2\omega)\Box \phi +2V(\phi)-\phi \frac{dV}{d\phi}=\kappa^2T
\end{equation}
In this theory we observe that both the matter and the scalar field act as sources for the metric, which means that both the matter and the scalar field generate the spacetime curvature. In fact, even in vacuum the scalar field curves the spacetime. According to the way we wrote the metric field equations, it is tempting to identify the Brans-Dicke field with a new matter field. However, since the Brans-Dicke scalar is sourced by the energy-momentum tensor (via its trace, which is a scalar magnitude constructed out of the sources of energy and momentum), we say that it is a gravitational field. Note, in this sense, that standard matter fields, such as a Dirac field coupled to electromagnetism $(i\gamma^\mu\partial_\mu-m)\psi=e\gamma^\mu A_\mu\psi$, do not couple to energy and momentum.

\section{Experimental determination of the gravity Lagrangian \label{sec:f(R)}}

Einstein's theory of general relativity (GR) represents one of the most impressive exercises of human intellect. As we have seen in previous sections, it implied a huge conceptual jump with respect to Newtonian gravity and, unlike the currently established standard model of elementary particles, no experiments were carried out to probe the structure of the theory. In spite of that, to date the theory has successfully passed all precision experimental tests. Its predictions are in agreement with experiments in scales that range from millimeters to astronomical units, scales in which weak and strong field phenomena can be observed \cite{Will05}. The theory is so successful in those regimes and scales that it is generally accepted that it should also work at larger and shorter scales, and at weaker and stronger regimes. \\
This extrapolation is, however, forcing us today to draw a picture of the universe that is not yet supported by other independent observations. For instance, to explain the rotation curves of spiral galaxies, we must accept the existence of vast amounts of unseen matter surrounding those galaxies. Additionally, to explain the luminosity-distance relation of distant type Ia supernovae and some properties of the  distribution of matter and radiation at large scales, we must accept the existence of yet another source of energy with repulsive gravitational properties (see \cite{CSTs}, \cite{Pad03}, \cite{PeRa03} for recent reviews on dark energy). Together those unseen (or dark) sources of matter and energy are found to make up to $96\%$ of the total energy of the observable universe! This huge discrepancy between the gravitationally estimated amounts of matter and energy and the direct measurements via electromagnetic radiation motivates the search for alternative theories of gravity which can account for the large scale dynamics and structure without the need for dark matter and/or dark energy. \\
In this sense, there has been an enormous international effort in the last years to determine whether the gravity Lagrangian could depart from Einstein's one at cosmic scales in a way compatible with the cosmological observations that support the cosmic speedup. In particular, many authors have investigated the consequences of promoting the Hilbert-Einstein Lagrangian to an arbitrary function $f(R)$ of the scalar curvature (see \cite{Olmo11a}, \cite{DFTs10}, \cite{SoFa08}, \cite{Capo-Mauro} for recent reviews). In this section we will show that the dynamics of the solar system can be used to set important constraints on the form of the function $f(R)$.

\subsection{Field equations of $f(R)$ theories.}

The action that defines $f(R)$ theories has the generic form
\begin{equation}\label{eq:def-f(R)}
S=\frac{1}{2\kappa ^2}\int d^4 x\sqrt{-g}f(R)+S_m[g_{\mu\nu},\psi_m] \ ,
\end{equation}
where $\kappa^2=8\pi G$, and we use the same notation introduced in previous sections.
Variation of (\ref{eq:def-f(R)}) leads to the following field
equations for the metric
\begin{equation}\label{eq:f-var}
f_R R_{\mu\nu}-\frac{1}{2}fg_{\mu\nu}-
\nabla_{\mu}\nabla_{\nu}f_R+g_{\mu\nu}\Box f_R=\kappa ^2T_{\mu
\nu }
\end{equation}
where $f_R\equiv df/dR$. According to (\ref{eq:f-var}), we
see that, in general, the metric satisfies a system of
fourth-order partial differential equations. The trace of
(\ref{eq:f-var}) takes the form
\begin{equation}\label{eq:trace-m}
3\Box f_R+Rf_R-2f=\kappa ^2T
\end{equation}
If we take $f(R)=R-2\Lambda$, (\ref{eq:f-var}) boils down to
\begin{equation}\label{eq:GR+L}
R_{\mu\nu}-\frac{1}{2}g_{\mu\nu}R=\kappa^2T_{\mu\nu}-\Lambda g_{\mu\nu} \ ,
\end{equation}
which represents GR with a cosmological constant. This is the only case in which an $f(R)$ Lagrangian yields second-order equations for the metric\footnote{This is so only if the connection is assumed to be the Levi-Civita connection of the metric (metric formalism). If the connection is regarded as independent of the metric, Palatini formalism, then $f(R)$ theories lead to second-order equations. This point will be explained in detail later on.}. \\

Let us now rewrite (\ref{eq:f-var}) in the form
\begin{equation}\label{eq:Gab-f(R)}
R_{\mu \nu }-\frac{1}{2}g_{\mu \nu }R=
\frac{\kappa^2}{f_R}T_{\mu\nu}-\frac{1}{2f_R}g_{\mu\nu}[Rf_R-f]+\frac{1}{f_R}\left[\nabla_\mu\nabla_\nu f_R-g_{\mu\nu}\Box f_R\right]
\end{equation}
The right hand side of this equation can now be seen as the source
terms for the metric. This equation, therefore, tells us that the
metric is generated by the matter and by terms related to the
scalar curvature. If we now wonder about what generates the scalar
curvature, the answer is in (\ref{eq:trace-m}). That expression
says that the scalar curvature satisfies a second-order
differential equation with the trace $T$ of the energy-momentum
tensor of the matter and other curvature terms acting as sources.
We have thus clarified the role of the higher-order derivative
terms present in (\ref{eq:f-var}). The scalar curvature is now
a dynamical entity which helps generate the space-time metric and
whose dynamics is determined by (\ref{eq:trace-m}).

At this point one should have noted the essential
difference between a generic $f(R)$ theory and GR. In GR the only
dynamical field is the metric and its form is fully characterized
by the matter distribution through the equations
$G_{\mu\nu}=\kappa^2T_{\mu\nu}$, where $G_{\mu\nu}\equiv R_{\mu\nu}-\frac{1}{2}g_{\mu\nu}R$. The scalar curvature is also
determined by the local matter distribution but through an
algebraic equation, namely, $R=-\kappa^2T$. In the $f(R)$ case
both $g_{\mu\nu}$ and $R$ are dynamical fields, i.e., they are
governed by differential equations. Furthermore,  the scalar
curvature $R$, which can obviously be expressed in terms of the
metric and its derivatives, now plays a non-trivial role in the determination of the metric itself.

The physical interpretation given above puts forward the central
and active role played by the scalar curvature in the field
equations of $f(R)$ theories. However, (\ref{eq:trace-m})
suggests that the actual dynamical entity is $f_R$ rather than
$R$ itself. This is so because, besides the metric, $f_R$ is the
only object acted on by differential operators in the field
equations. Motivated by this, we can introduce the following
alternative notation
\begin{eqnarray}\label{eq:phi=f_R}
\phi&\equiv& f_R\\
V(\phi)&\equiv& R(\phi)f_R-f(R(\phi)) \label{eq:V=rf_R-f}
\end{eqnarray}
and rewrite eqs. (\ref{eq:Gab-f(R)}) and (\ref{eq:trace-m}) in the same form as (\ref{eq:Gab-ST}) and (\ref{eq:phi-ST}) with the choice $w=0$.
This slight change of notation helps us identify the $f(R)$ theory in metric formalism with a scalar-tensor Brans-Dicke theory with parameter $\omega=0$ and
non-trivial potential $V(\phi)$, whose action was given in (\ref{eq:ST}).
In terms of this scalar-tensor representation our interpretation
of the field equations of $f(R)$ theories is obvious, since both
the matter and the scalar field help generate the metric. The
scalar field is a dynamical object influenced by the matter and by
self-interactions according to (\ref{eq:phi-ST}).

\subsection{Spherically symmetric systems}\label{sec:calculations}

 A complete description of a physical system must take into account not only the system but also its interaction with the environment. In this sense, any physical system is surrounded by the rest of the universe. The relation of the local system with the rest of the universe manifests itself in a set of boundary conditions. In our case, according to (\ref{eq:trace-m}) and (\ref{eq:Gab-f(R)}), the metric and the function $f_R$ (or, equivalently, $R$ or $\phi$) are subject to boundary conditions, since they are dynamical fields (they are governed by differential equations). The boundary conditions for the metric can be trivialized by a suitable choice of coordinates. In other words, we can make the metric Minkowskian in the asymptotic region and fix its first derivatives to zero (see chapter 4 of \cite{Will93} for details). The function $f_R$, on the other hand, should tend to the cosmic value $f_{R_c}$ as we move away from the local system. The precise value of $f_{R_c}$ is obtained by solving the equations of motion for the corresponding cosmology. According to this, the local system will interact with the asymptotic (or background) cosmology via the boundary value $f_{R_c}$ and its cosmic-time derivative. Since the cosmic time-scale is  much larger than the typical time-scale of local systems (billions of years versus years), we can assume an adiabatic interaction between the local system and the background cosmology. We can thus neglect terms such as $\dot{f}_{R_c}$, where dot denotes derivative with respect to the cosmic time.\\
The problem of finding solutions for the local system, therefore, reduces to solving (\ref{eq:Gab-f(R)}) expanding about the Minkowski metric in the asymptotic region\footnote{Note that the expansion about the Minkowski metric does not imply the existence of global Minkowskian solutions. As we will see, the general solutions to our problem turn out to be asymptotically de Sitter spacetimes.}, and (\ref{eq:trace-m}) tending  to
\begin{equation}\label{eq:trace-cosmo}
3\Box_c f_{R_c}+R_cf_{R_c}-2f(R_c)=\kappa^2 T_c
\end{equation}
where the subscript $c$ denotes cosmic value, far away from the system. In particular, if we consider a weakly gravitating local system, we can take $f_R=f_{R_c}+\varphi(x)$ and $g_{\mu\nu}=\eta_{\mu\nu}+h_{\mu\nu}$, with $|\varphi|\ll|f_{R_c}|$ and $|h_{\mu\nu}|\ll 1$ satisfying  $\varphi\to 0$ and $h_{\mu\nu}\to 0$ in the asymptotic region. Note that should the local system represent a strongly gravitating system such as a neutron star or a black hole, the perturbative expansion would not be sufficient everywhere. In such cases, the perturbative approach would only be valid in the far region. Nonetheless, the decomposition $f_R=f_{R_c}+\varphi(x)$ is still very useful because the equation for the local deviation $\varphi(x)$ can be written as
\begin{equation}\label{eq:trace-local}
3\Box \varphi +W(f_{R_c}+\varphi)-W(f_{R_c})=\kappa^2T,
\end{equation}
where $T$ represents the trace of the local sources, we have defined $W(f_R)\equiv R(f_R)f_R-2f(R[f_R])$, and $W(f_{R_c}) $ is a slowly changing constant within the adiabatic approximation. In this case, $\varphi$ needs not be small compared to $f_{R_c}$ everywhere, only in the asymptotic regions.

\subsubsection{Spherically symmetric solutions}

Let us define the line element\footnote{As pointed out in \cite{Olmo07}, solar system tests are conventionally described in isotropic coordinates rather than on Schwarzschild-like coordinates. This justifies our coordinate choice in (\ref{eq:ds2-iso}).}  \cite{Olmo07}
\begin{equation}\label{eq:ds2-iso}
ds^2=-A({r})e^{2\psi({r})}dt^2+\frac{1}{A({r})}\left(d{r}^2+{r}^2d\Omega^2\right),
\end{equation}
which, assuming a perfect fluid for the sources, leads to the following field equations
\begin{eqnarray}\label{eq:Ar-iso}
A_{rr}+A_r\left[\frac{2}{r}-\frac{5}{4}\frac{A_r}{A}\right]&=&\frac{\kappa^2\rho}{{f_R}}+\frac{R
{f_R}-f(R)}{2{f_R}}+ \frac{A}{{f_R}}\left[{f_R}_{rr}+{f_R}_r\left(\frac{2}{r}-\frac{A_r}{2A}\right)\right]\\
A\psi_r\left[\frac{2}{r}+\frac{{f_R}_r}{{f_R}}-\frac{A_r}{A}\right]-\frac{A^2_r}{4A}&=&\frac{\kappa^2P}{{f_R}}-\frac{R{f_R}-f(R)}{2{f_R}}- A\frac{{f_R}_r}{{f_R}}\left[\frac{2}{r}-\frac{A_r}{2A}\right]\label{eq:psir-iso}
\end{eqnarray}
where ${f_R}={f_R}_c+\varphi$, and the subscripts {\it r} in
$\psi_r, \ {f_R}_r, \ {f_R}_{rr}, \ M_r$ denote derivation with respect to the
radial coordinate. Note also that ${f_R}_r=\varphi_r$ and ${f_R}_{rr}=\varphi_{rr}$.
The equation for $\varphi$ is, according to (\ref{eq:trace-local}) and (\ref{eq:ds2-iso}),
\begin{equation} \label{eq:frr-iso}
A\varphi_{rr}=-A\left(\frac{2}{r}+\psi_r\right)\varphi_r -\frac{W({f_R}_c+\varphi)-W({f_R}_c)}{3}+\frac{\kappa^2}{3}(3P-\rho)
\end{equation}
Equations (\ref{eq:Ar-iso}), (\ref{eq:psir-iso}),  and (\ref{eq:frr-iso}) can be used to work out the metric of any spherically symmetric system subject to the asymptotic boundary conditions discussed above. For weak sources, such as non-relativistic stars like the sun, it is convenient to expand them assuming $|\varphi|\ll {f_R}_c$ and $A=1-2M(r)/r$, with $2M(r)/r\ll1$. The result is
\begin{eqnarray}\label{eq:Ar-iso-lin}
-\frac{2}{r}M_{rr}(r)&=&\frac{\kappa^2\rho}{{f_R}_c}+V_c+\frac{1}{{f_R}_c}\left[\varphi_{rr}+\frac{2}{r}\varphi_r\right]\\
\frac{2}{r}\left[\psi_r+\frac{\varphi_r}{{f_R}_c}\right]&=&\frac{\kappa^2}{{f_R}_c}P-V_c\label{eq:psir-iso-lin}\\
\varphi_{rr}+\frac{2}{r}\varphi_r-m_c^2\varphi&=&\frac{
\kappa^2}{3}(3P-\rho)
\label{eq:frr-iso-lin}
\end{eqnarray}
where we have defined
\begin{equation}
V_c\equiv \left.\frac{R{f_R}-f}{2{f_R}}\right|_{R_c} \  \mbox{ and } \
m^2_c\equiv\left.\frac{{f_R}-R f_{RR}}{3f_{RR}}\right|_{R_c} \ . \label{eq:mass}
\end{equation}
This expression for $m^2_c$ was first found in \cite{Olmo05a,Olmo05b} within the scalar-tensor approach. It was found there that $m^2_c>0$ is needed to have a well-behaved (non-oscillating) Newtonian limit. This expression and the conclusion $m^2_c>0$ were also reached in \cite{F-N05} by studying the stability of de Sitter space. The same expression has been rediscovered later several times.\\
Outside of the sources, the solutions of (\ref{eq:Ar-iso-lin}), (\ref{eq:psir-iso-lin}) and (\ref{eq:frr-iso-lin}) lead to
\begin{eqnarray}
\varphi(r)&=&\frac{C_1}{r}e^{-m_c r}\label{eq:phir-iso-sol}\\
A(r)&=&1-\frac{C_2}{r}\left(1-\frac{C_1}{C_2 {f_R}_c}e^{-m_c r}\right)+\frac{V_c}{6}r^2\label{eq:Ar-iso-sol}\\
A(r)e^{2\psi}&=& 1-\frac{C_2}{r}\left(1+\frac{C_1}{C_2 {f_R}_c}e^{-m_c r}\right)-\frac{V_c}{3}r^2\label{eq:psir-iso-sol}
\end{eqnarray}
where an integration constant $\psi_0$ has been absorbed in a redefinition of the time coordinate. The above solutions coincide, as expected, with those found in \cite{Olmo05a,Olmo05b} for the Newtonian and post-Newtonian limits using the scalar-tensor representation and standard gauge choices in Cartesian coordinates. Comparing our solutions with those, we identify
\begin{equation}
C_2\equiv \frac{\kappa^2}{4\pi {f_R}_c}M_\odot \ \mbox{  and  } \ \frac{C_1}{{f_R}_cC_2}\equiv\frac{1}{3} \label{eq:C1}
\end{equation}
where $M_\odot=\int d^3x \rho(x)$. The line element (\ref{eq:ds2-iso}) can thus be written as
\begin{equation}\label{eq:ds2-iso-sol}
ds^2=-\left(1-\frac{2G M_\odot}{r}-\frac{V_c}{3}r^2\right)dt^2+\left(1+\frac{2G\gamma M_\odot}{r}-\frac{V_c}{6}r^2\right)(dr^2+r^2d\Omega^2)
\end{equation}
where we have defined the effective Newton's constant and post-Newtonian parameter $\gamma$ as
\begin{equation}\label{eq:G}
G=\frac{\kappa^2}{8\pi {f_R}_c}\left(1+\frac{e^{-m_cr}}{3}\right) \ \mbox{ and } \ \gamma=\frac{3-e^{-m_cr}}{3+e^{-m_cr}}
\end{equation}
respectively. This completes the lowest-order solution in isotropic coordinates.

\subsubsection{The gravity Lagrangian according to solar system experiments}

From the definitions of Eq.(\ref{eq:G}) we see that the parameters $G$ and $\gamma$ that characterize the linearized metric depend on the effective mass $m_c$ (or inverse length scale $\lambda_{m_c}\equiv m_c^{-1}$). Newton's constant, in addition, also depends on ${f_{R}}_c$. Since the value of the background cosmic curvature $R_c$ changes with the cosmic expansion, it follows that ${f_{R}}_c$ and $m_c$ must also change. The variation in time of ${f_{R}}_c$ induces a time variation in the effective Newton's constant which is just the well-known time dependence that exists in Brans-Dicke theories. The length scale $\lambda_{m_c}$, characteristic of $f(R)$ theories, does not appear in the original Brans-Dicke theories because in the latter the scalar potential was assumed to vanish,  $V(\phi)\equiv 0$, in contrast with (\ref{eq:V=rf_R-f}), which implies an infinite interaction range ($m_c=0 \ \rightarrow \ \lambda_{m_c}=\infty$).

In order to have agreement with the observed properties of the solar system, the Lagrangian $f(R)$ must satisfy certain basic constraints. These constraints will be very useful to determine the viability of some families of models proposed to explain the cosmic speedup. A very representative family of such models, which do exhibit self-accelerating late-time cosmic solutions, is given by $f(R)=R-R_0^{n+1}/R^n$, where $R_0$ is a very low curvature scale that sets the scale at which the model departs from GR, and $n$ is assumed positive. At curvatures higher than $R_0$, the theory is expected to behave like GR while at late times, when the cosmic density decays due to the expansion and approaches the scale $R_0$, the modified dynamics becomes important and could explain the observed speedup. \\
In viable theories, the effective cosmological constant $V_c$ must be negligible. Most importantly, the interaction range $\lambda_{m_c}$ must be shorter than a few millimeters because such Yukawa-type corrections to the Newtonian potential have not been observed, and observations indicate that the parameter $\gamma$ is very close to unity. This last constraint can be expressed as $(L_S^2/\lambda_{m_c}^2)\gg 1$, where $L_S$ represents a (relatively short) length scale that can range from meters to planetary scales, depending on the particular test used to verify the theory.  In terms of the Lagrangian, this constraint takes the form
\begin{equation}\label{eq:c1}
\left.\frac{{f_R}-R f_{RR}}{3f_{RR}}\right|_{R_c}L^2_S\gg 1 \ .
\end{equation}
A qualitative analysis of this constraint can be used to argue that, in general, $f(R)$ theories with terms
that become dominant at low cosmic curvatures, such as the models $f(R)=R-R_0^{n+1}/R^n$, are not viable
theories in solar system scales  and, therefore, cannot represent
an acceptable mechanism for the cosmic expansion.\\
Roughly speaking, eq.(\ref{eq:c1}) says that the smaller the term $f_{RR}(R_c)$,
with $f_{RR}(R_c)>0$ to guarantee $m_\varphi ^2>0$, the heavier the
scalar field. In other words, the smaller $f_{RR}(R_c)$, the shorter the interaction range of the field.
In the limit $f_{RR}(R_c)\to 0$, corresponding to GR, the scalar
interaction is completely suppressed. Thus, if the nonlinearity of
the gravity Lagrangian had become dominant in the last few
billions of years (at low cosmic curvatures), the scalar field
interaction range $\lambda_{m_c}$ would have increased accordingly. In
consequence, gravitating systems such as the solar system,
globular clusters, galaxies,\ldots would have experienced (or will
experience) observable changes in their gravitational dynamics.
Since there is no experimental evidence supporting such a change\footnote{As an example, note that the fifth-force effects of the Yukawa-type correction introduced by the scalar degree of freedom would have an effect on stellar structures and their evolution, which would lead to incompatibilities with current observations.}
and all currently available solar system gravitational experiments
are compatible with GR, it seems unlikely that the nonlinear
corrections may be dominant at the current epoch.

Let us now analyze in detail the constraint given in
eq.(\ref{eq:c1}). That equation can be rewritten as follows
\begin{equation}\label{eq:c11}
R_c\left[\left.\frac{f_R}{R f_{RR}}\right|_{R_c}-1\right]L_S^2\gg 1
\end{equation}
We are interested in the form of the Lagrangian at intermediate
and low cosmic curvatures, i.e., from the matter dominated to the
vacuum dominated eras. We shall now demand that the interaction
range of the scalar field remains as short as today or decreases
with time so as to avoid dramatic modifications of the
gravitational dynamics in post-Newtonian systems with the cosmic
expansion. This can be implemented imposing
\begin{equation}\label{eq:difeq-0}
\left[\frac{f_R}{Rf_{RR}}-1\right]\ge \frac{1}{l^2R}
\end{equation}
as $R\to 0$, where $l^2\ll L_S^2$ represents a bound to the
current interaction range of the scalar field. Thus,
eq.(\ref{eq:difeq-0}) means that the interaction range of the
field must decrease or remain short, $\sim l^2$, with the
expansion of the universe. Manipulating this expression, we obtain
\begin{equation}\label{eq:difeq-1}
\frac{d\log[f_R]}{dR}\le \frac{l^2}{1+l^2R}
\end{equation}
which can be integrated twice to give the following inequality
\begin{equation}\label{eq:fR-0}
f(R)\le A+B\left(R+\frac{l^2R^2}{2}\right)
\end{equation}
where $B$ is a positive constant, which can be set to unity
without loss of generality. Since $f_R$ and $f_{RR}$ are positive, the
Lagrangian is also bounded from below, i.e., $f(R)\ge A$. In
addition, according to the cosmological data, $A \equiv -2\Lambda$
must be of order a cosmological constant $2\Lambda \sim 10^{-53} $
m$^2$. We thus conclude that the gravity Lagrangian at
intermediate and low scalar curvatures is bounded by
\begin{equation}\label{eq:fR-1}
-2\Lambda \le f(R)\le R-2\Lambda +\frac{l^2R^2}{2}
\end{equation}
This result shows that a Lagrangian with nonlinear terms that grow
with the cosmic expansion is not compatible with the current solar
system gravitational tests, such as we argued above. Therefore,
those theories cannot represent a valid mechanism to justify the
observed cosmic speed-up. Additionally, our analysis has provided
an empirical procedure to determine the form of the gravitational Lagrangian. The function $f(R)$ found here nicely recovers Einstein's gravity at low curvatures but allows for some quadratic corrections at higher curvatures, which is of interest in studies of the very early Universe.

\section{Quantum gravity phenomenology and the early universe}

The extrapolation of the dynamics of GR to the very strong field regime indicates that the Universe began at a singularity and that the death of a sufficiently massive star unavoidably leads to the formation of a black hole or a naked singularity. The existence of space-time singularities is one of the most impressive predictions of GR. This prediction, however, also represents the end of the theory, because the absence of a well-defined geometry implies the absence of physical laws and lack of predictability \cite{Novello-2008,Hawking-1975}. For this reason, it is generally accepted that the dynamics of GR must be changed at some point to avoid these problems. A widespread belief is that at sufficiently high energies the gravitational field must exhibit quantum properties that alter the dynamics and prevent the formation of singularities. However, a completely satisfactory quantum theory of gravity is not yet available. To make some progress in the qualitative understanding of how quantum gravity may affect the dynamics of the Universe near the big bang, in this section we show how certain modifications of GR may be able to capture some aspects of the expected phenomenology of quantum gravity \cite{Olmo11b}.\\
We begin by noting that Newton's and Planck's constants may be combined with the speed of  light to generate a length $l_P=\sqrt{\hbar G/c^3}$,  which is known as the Planck length. The Planck length is usually interpreted as the scale at which quantum gravitational phenomena should play a non-negligible role. However, since lengths are not relativistic invariants, the existence of the Planck length raises doubts about the nature of the reference frame in which it should be measured and about the limits of validity of special relativity itself. This poses the following question: can we combine in the same framework the speed of light and the Planck length in such a way that both quantities appear as universal invariants to all observers? The solution to this problem will give us the key to consider quantum gravitational phenomena from a modified gravity perspective.

\subsection{Palatini approach to modified gravity}

To combine in the same framework the speed of light and the Planck length in a way that preserves the invariant and universal
nature of both quantities, we first note that though $c^2$ has dimensions of squared velocity it represents a 4-dimensional Lorentz scalar rather than the squared of a privileged 3-velocity. Similarly, we may see $l_P^2$ as a 4-d invariant with dimensions of length squared that needs not be related with any privileged 3-length. Because of dimensional compatibility with a curvature, the invariant $l_P^2$ could be introduced in the theory via the gravitational sector by considering departures from GR at the Planck scale motivated by quantum effects. However, the situation is not as simple as it may seem at first. In fact, an action like the one we obtained in the last section\footnote{Restoring missing factors of $c$ in (\ref{eq:def-f(R)}), we find that $\frac{1}{16\pi G}=\frac{\hbar}{16\pi l_P^2}$ and, therefore, $\kappa^2=8\pi l_P^2/\hbar$.},
\begin{equation}\label{eq:f(R)}
S[g_{\mu\nu},\psi]=\frac{\hbar}{16\pi l_P^2}\int d^4x \sqrt{-g}\left[R+l_P^2 R^2\right]+S_m[g_{\mu\nu},\psi] \ ,
\end{equation}
where $S_m[g_{\mu\nu},\psi]$ represents the matter sector, contains the scale $l_P^{2}$ but not in the invariant form that we wished. The reason is that the field equations that follow from (\ref{eq:f(R)}) are equivalent to those of the following scalar-tensor theory
\begin{equation}\label{eq:f(R)-met}
S[g_{\mu\nu},\varphi,\psi]=\frac{\hbar}{16\pi l_P^2}\int d^4x \sqrt{-g}\left[(1+\varphi) R-\frac{1}{4l_P^2}\varphi^2\right]+S_m[g_{\mu\nu},\psi] \ ,
\end{equation}
which given the identification $\phi=1+\varphi$ coincides with the case $w=0$ of Brans-Dicke theory with a non-zero potential $V(\phi)=\frac{1}{4l_P^2}(\phi-1)^2$. As is well-known and was explicitly shown in Section \ref{sec:f(R)}, in Brans-Dicke theory the observed Newton's constant is promoted to the status of field, $G_{eff}\sim G/\phi$. The scalar field allows the effective Newton's constant $G_{eff}$ to dynamically change in time and in space. As a result the corresponding effective Planck length, $\tilde{l}_P^2=l_P^2/\phi$, would also vary in space and time. This is quite different from the assumed constancy and universality of the speed of light in special relativity, which is  implicit in our construction of the total action. In fact, our action has been constructed assuming the Einstein equivalence principle (EEP),
whose validity guarantees that the observed speed of light is a true constant and universal invariant, not a field\footnote{If the Einstein equivalence principle is true, then all the coupling constants of the standard model are constants, not fields \cite{Will05}.}
like in varying speed of light theories \cite{VSL} (recall also that Nordstr\"{o}m's first scalar theory was motivated by the constancy of the speed of light). The situation does not improve if we introduce higher curvature invariants in (\ref{eq:f(R)-met}).  We thus see that the introduction of the Planck length in the gravitational sector in the form of a universal constant like the speed of light is not a trivial issue. The introduction of curvature invariants suppressed by powers of $R_P=1/l_P^2$ unavoidably generates new degrees of freedom which turn Newton's constant into a dynamical field.

\indent Is it then possible to modify the gravity Lagrangian adding Planck-scale corrected terms without turning Newton's constant into a dynamical field? The answer to this question is in the affirmative. One must first note that metricity and affinity are a priori logically independent concepts \cite{Zanelli}. If we construct the theory {\it \`{a} la Palatini}, that is in terms of a connection not a priori constrained to be given by the Christoffel symbols, then the resulting  equations do not necessarily contain new dynamical degrees of freedom (as compared to GR), and the Planck length may remain space-time independent in much the same way as the speed of light and the coupling constants of the standard model, as required by the {EEP}. A natural alternative, therefore, seems to be to consider (\ref{eq:f(R)}) in the Palatini formulation. The field equations that follow from (\ref{eq:f(R)}) when metric and connection are varied independently are \cite{Olmo11a}
\begin{eqnarray}
f_R R_{\mu\nu}(\Gamma)-\frac{1}{2}f g_{\mu\nu}&=&\kappa^2 T_{\mu\nu} \label{eq:metric}\\
\nabla_\alpha\left(\sqrt{-g}f_R g^{\beta\gamma}\right)&=&0 \ , \label{eq:connection}
\end{eqnarray}
where $f=R+R^2/R_P$, $f_R\equiv \partial_R f=1+2R/R_P$, $R_P=1/l_P^2$, and $\kappa^2=8\pi l_P^2/\hbar$. The connection equation (\ref{eq:connection}) can be easily solved after noticing that the trace of (\ref{eq:metric}) with $g^{\mu\nu}$,
\begin{equation}\label{eq:trace-Pal}
R f_R-2f=\kappa^2T \ ,
\end{equation}
represents an algebraic relation between $R\equiv g^{\mu\nu}R_{\mu\nu}(\Gamma)$ and $T$, which generically implies that $R=R(T)$ and hence $f_R=f_R(R(T))$ [from now on we denote $f_R(T)\equiv f_R(R(T))$]. For the particular Lagrangian (\ref{eq:f(R)}), we find that $R=-\kappa^2T$, like in GR. This relation implies that (\ref{eq:connection}) is just a first order equation for the connection that involves the matter, via the trace $T$, and the metric. The connection turns out to be the Levi-Civita connection of an auxiliary metric,
\begin{equation}
\Gamma^\alpha_{\mu\nu}=\frac{h^{\alpha\beta}}{2}\left(\partial_\mu h_{\beta\nu}+\partial_\nu h_{\beta\mu}-\partial_\beta h_{\mu\nu}\right) \ ,
\end{equation}
which is conformally related with the physical metric, $h_{\mu\nu}= f_R(T) g_{\mu\nu}$. Now that the connection has been expressed in terms of $h_{\mu\nu}$, we can rewrite (\ref{eq:metric}) as follows
\begin{equation}\label{eq:Gmn-pal}
G_{\mu\nu}(h)=\frac{\kappa^2}{f_R(T)} T_{\mu\nu}-\Lambda(T)h_{\mu\nu}
\end{equation}
where $\Lambda(T)\equiv (R f_R-f)/(2f_R^2)=(\kappa^2 T)^2/R_P$, and looks like Einstein's theory for the metric $h_{\mu\nu}$ with a slightly modified source. This set of equations can also be written in terms of the physical metric $g_{\mu\nu}$ as follows
\begin{eqnarray}\label{eq:Gab-fPal}
R_{\mu \nu }(g)-\frac{1}{2}g_{\mu \nu }R(g)&=&\frac{\kappa
^2}{f_R}T_{\mu \nu }-\frac{Rf_R-f}{2f_R}g_{\mu \nu
}-\frac{3}{2(f_R)^2}\left[\partial_\mu f_R\partial_\nu
f_R-\frac{1}{2}g_{\mu \nu }(\partial f_R)^2\right]+ \nonumber \\ & &\frac{1}{f_R}\left[\nabla_\mu \nabla_\nu f_R-g_{\mu \nu }\Box
f_R\right] \ .
\end{eqnarray}
In this last representation, one can use the notation introduced in (\ref{eq:phi=f_R}) and (\ref{eq:V=rf_R-f}) to show that these field equations coincide with those of a Brans-Dicke theory with parameter $w=-3/2$ (see eq.(\ref{eq:Gab-ST})). Note that all the functions $f(R)$, $R$, and $f_R$ that appear on the right hand side of (\ref{eq:Gab-fPal}) are functions of the trace $T$. This means that the modified dynamics of (\ref{eq:Gab-fPal}) is due to the new matter terms induced by the trace $T$ of the matter, not to the presence of new dynamical degrees of freedom. This also guarantees that, unlike for the $w\neq -3/2$ Brans-Dicke theories, for the $w=-3/2$ theory Newton's constant is indeed a constant. \\
From the structure of the field equations (\ref{eq:Gmn-pal}) and (\ref{eq:Gab-fPal}), and the relation $g_{\mu\nu}=(1/f_R) h_{\mu\nu}$, it follows that $g_{\mu\nu}$ is affected by the matter-energy in two different ways. The first contribution corresponds to the cumulative effects of matter, and the second contribution is due to the dependence on the local density distributions of energy and momentum. This can be seen by noticing that the structure of the equations (\ref{eq:Gmn-pal}) that determine $h_{\mu\nu}$ is similar to that of GR, which implies that $h_{\mu\nu}$ is determined by integrating over all the sources (gravity as a cumulative effect).  Besides that, $g_{\mu\nu}$ is also affected by the local sources through the factor $f_R(T)$. To illustrate this point, consider a region of the spacetime containing a total mass $M$ and filled with sources of low energy-density as compared to the Planck scale ($|\kappa^2 T/R_P|\ll 1$). For the quadratic model $f(R)=R+R^2/R_P$, in this region (\ref{eq:Gmn-pal}) boils down to $G_{\mu\nu}(h)=\kappa^2T_{\mu\nu}+O(\kappa^2T/R_P)$, and $h_{\mu\nu}\approx (1+O(\kappa^2T/R_P))g_{\mu\nu}$, which implies that the GR solution is a very good approximation. This confirms that $h_{\mu\nu}$ is determined by an integration over the sources, like in GR. Now, if this region is traversed by a particle of mass $m\ll M$ but with a non-negligible ratio $\kappa^2T/R_P$, then the contribution of this particle to $h_{\mu\nu}$ can be neglected, but its effect on $g_{\mu\nu}$ via de factor $f_R=1-\kappa^2T/R_P$ on the region that supports the particle (its classical trajectory) is important. This phenomenon is analogous to that described in the so-called Rainbow Gravity \cite{Magueijo:2002xx}, an approach to the phenomenology of quantum gravity based on a non-linear implementation of the Lorentz group to allow for the coexistence of a constant speed of light and a maximum energy scale (the flat space version of that theory is known as Doubly Special Relativity \cite{DSR1,DSR2a,DSR2b,AmelinoCamelia:2009pg}). In Rainbow Gravity, particles of different energies (energy-densities in our case) perceive different metrics.

\subsection{The early-time cosmology of Palatini $f(R)$ models.}

The quadratic Palatini model introduced above turns out to be virtually indistinguishable from GR at energy densities well below the Planck scale. It is thus natural to ask if this theory presents any particularly interesting feature at Planck scale densities. A natural context where this question can be explored is found in the very early universe, when the matter energy-density tends to infinity as we approach the big bang. \\
 In a spatially flat, homogeneous, and isotropic universe, with line element $ds^2=-dt^2+a^2(t)d\vec{x}^2$, filled with a perfect fluid with constant equation of state $P=w\rho$ and density $\rho$, the Hubble function that follows from (\ref{eq:Gab-fPal}) (or (\ref{eq:Gmn-pal})) takes the form
\begin{equation}\label{eq:Hubble-iso}
H^2=\frac{1}{6f_R}\frac{\left[f+(1+3w)\kappa^2\rho\right]}{\left[1+\frac{3}{2}\Delta\right]^2}  \ ,
\end{equation}
where $H=\dot{a}/a$, and ${\Delta}=-(1+w)\rho\partial_\rho f_R/f_R=(1+w)(1-3w)\kappa^2\rho  f_{RR}/(f_R(Rf_{RR}-f_R))$. In GR, (\ref{eq:Hubble-iso}) boils down to $H^2=\kappa^2\rho/3$. Since the matter conservation equation for constant $w$ leads to $\rho=\rho_0/a^{3(1+w)}$, in GR we find that $a(t)=a_0 t^{\frac{2}{3(1+w)}}$, where $\rho_0$ and $a_0$ are constants. This result indicates that if the universe is dominated by a matter source with $w>-1$, then at $t=0$ the universe has zero physical volume, the density is infinite, and all curvature scalars blow up, which indicates the existence of a big bang singularity. The quadratic Palatini model introduced above, however, can avoid this situation. For that model, (\ref{eq:Hubble-iso}) becomes \cite{BOSA09a,BOSA09b,Olmo10}
\begin{equation}\label{eq:H-R2}
H^2=\frac{\kappa^3\rho}{3}\frac{\left(1+\frac{2R}{R_P}\right)\left(1+\frac{1-3w}{2}\frac{R}{R_P}\right)}{\left[1-(1+3w)\frac{R}{R_P}\right]^2} \ .
\end{equation}
This expression recovers the linear dependence on $\rho$ of GR in the limit $|R/R_P|\ll 1$. However, if $R$ reaches the value $R_b=-R_P/2$, then $H^2$ vanishes and the expansion factor $a(t)$ reaches a minimum. This occurs for $w>1/3$ if $R_P>0$ and for $w<1/3$ if $R_P<0$. The existence of a non-zero minimum for the expansion factor implies that the big bang singularity is avoided. The avoidance of the big bang singularity indicates that the time coordinate can be extended backwards in time beyond the instant $t=0$. This means that in the past the universe was in a contracting phase which reached a minimum and bounced off to the expanding phase that we find in GR.

We mentioned at the beginning of this section that the avoidance of the big bang singularity is a basic requirement for any acceptable quantum theory of gravity. Our procedure to construct a quantum-corrected theory of gravity in which the Planck length were a universal invariant similar to the speed of light has led us to a cosmological model which replaces the big bang by a cosmic bounce. To obtain this result, it has been necessary to resort to the Palatini formulation of the theory. In this sense, it is important to note that the metric formulation of the quadratic curvature model discussed here, besides turning the Planck length into a dynamical field, is unable to avoid the big bang singularity. In fact, in metric formalism, all quadratic models of the form $R+(a R^2+bR_{\mu\nu}R^{\mu\nu})/R_P$ that at late times tend to a standard Friedmann-Robertson-Walker cosmology begin with a big bang singularity. Palatini theories, therefore, appear as a potentially interesting framework to discuss quantum gravity phenomenology.

\subsection{A Palatini action for loop quantum cosmoloy}

\begin{figure}[htb]
\centering
\includegraphics[width=8cm]{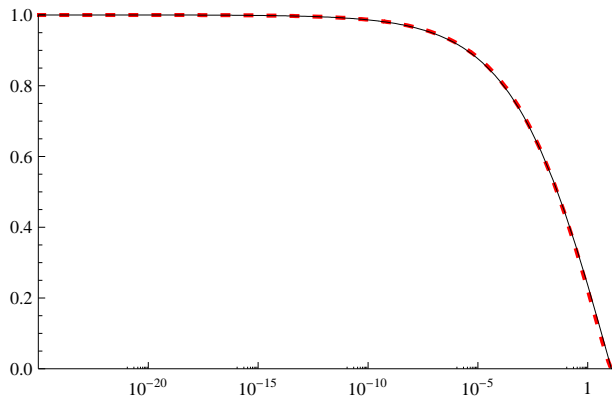}
\caption{Vertical axis: {$df/dR$} \ ; \ Horizontal axis: {$R/R_c$}. Comparison of the numerical solution with the interpolating function (\ref{eq:f-guess}). The dashed line represents the numerical curve.}\label{LQC-plot}
\end{figure}

Growing interest in the dynamics of the early-universe in Palatini theories has arisen, in part, from the observation that the effective equations of loop quantum cosmology (LQC) \cite{Bojo05,APS06a,APS06b,APS06c,Ash07,SKL07},
a Hamiltonian approach to quantum gravity based on the non-perturbative quantization techniques of loop quantum gravity \cite{Rovelli,Thiemann}, could be exactly reproduced by a Palatini $f(R)$ Lagrangian \cite{OS-2009}. In LQC, non-perturbative quantum gravity effects lead to the resolution of the big bang singularity by a quantum bounce without introducing any new degrees of freedom. Though fundamentally discrete, the theory admits a continuum description in terms of an effective Hamiltonian that in the case of a homogeneous and isotropic universe filled with a massless scalar field leads to the following modified Friedmann equation
\begin{equation}\label{eq:LQC}
3H^2={8\pi G}\rho\left(1-\frac{\rho}{\rho_{crit}}\right) \ ,
\end{equation}
where $\rho_{crit}\approx 0.41\rho_{Planck}$. At low densities, $\rho/\rho_{crit}\ll 1$, the background dynamics is the same as in GR, whereas at densities of order $\rho_{crit}$ the non-linear new matter contribution forces the vanishing of $H^2$ and hence a cosmic bounce. This singularity avoidance seems to be a generic feature of loop-quantized universes \cite{Param09}. \\
Palatini $f(R)$ theories share with LQC an interesting property: the modified dynamics that they generate is not the result of the existence of new dynamical degrees of freedom but rather it manifests itself by means of non-linear contributions produced by the matter sources, which contrasts with other approaches to quantum gravity and to modified gravity. This similarity makes it tempting to put into correspondence Eq.(\ref{eq:LQC}) with the corresponding $f(R)$ equation (\ref{eq:Gab-fPal}).
Taking into account the trace equation (\ref{eq:trace-Pal}), which for a massless scalar becomes $R f_R -2f=2\kappa^2\rho$ and implies that $\rho=\rho({R})$, one finds that a Palatini $f(R)$ theory able to reproduce the LQC dynamics (\ref{eq:LQC}) must satisfy the differential equation
\begin{equation}
f_{RR}=-f_R\left(\frac{A f_R -B}{2({R}f_R-3f)A+{R}B}\right)
\end{equation}
where $A=\sqrt{2({R}f_R-2f)(2{R}_c-[Rf_R-2f])}$, $B=2\sqrt{{R}_cf_R(2{R} f_R-3f)}$, and ${R}_c\equiv \kappa^2\rho_c$. If one imposes the boundary condition $\lim_{R\to 0} f_R\to 1$ at low curvatures, and $\ddot{a}_{LQC}=\ddot{a}_{Pal}$ (where $\ddot{a}$ represents the acceleration of the expansion factor) at $\rho=\rho_c$, the solution to this equation is unique.  The solution was found numerically \cite{OS-2009}, though the following function can be regarded as a very accurate approximation to the LQC dynamics from the GR regime to the non-perturbative bouncing region (see Fig.\ref{LQC-plot})
\begin{equation}\label{eq:f-guess}
\frac{df}{dR}=- \tanh \left(\frac{5}{103}\ln\left[\left(\frac{R}{12\mathcal{R}_c}\right)^2\right]\right)
\end{equation}
\\

This result is particularly important because it establishes a direct link between the Palatini approach to modified gravity and a cosmological model derived from non-perturbative quantization techniques.

\subsection{Beyond Palatini $f(R)$ theories.}

\begin{figure}[htb]	
\centering
\includegraphics[width=8cm]{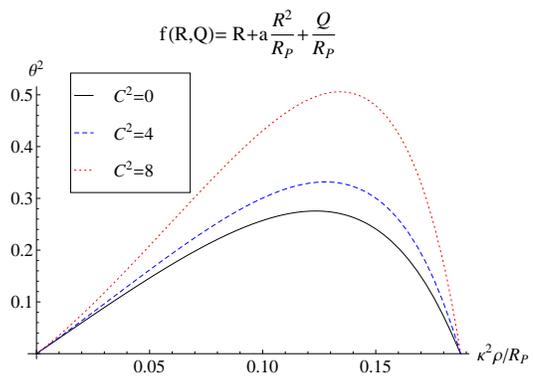}
\caption{Evolution of the expansion as a function of $\kappa^2\rho/R_P$ in radiation universes with low anisotropy, which is controlled by the combination $C^2=C_{12}^2+C_{23}^2+C_{31}^2$. The case with $C^2=0$ corresponds to the isotropic flat case, $\theta^2=9H^2$. }\label{fig:ExpanRad}
\end{figure}

Nordstr\"{o}m's second theory was a very interesting theoretical exercise that successfully allowed to implement the Einstein equivalence principle in a relativistic scalar theory. However, among other limitations, that theory did not predict any new gravitational effect for the electromagnetic field. In a sense, Palatini $f(R)$ theories suffer from this same limitation. Since their modified dynamics is due to new matter contributions that depend on the trace of the stress-energy tensor, for traceless fields such as a radiation fluid or the electromagnetic field, the theory does not predict any new effect. This drawback can be avoided if one adds to the Palatini Lagrangian a new piece dependent on the squared Ricci tensor, $R_{\mu\nu}R^{\mu\nu}$, where we assume $R_{\mu\nu}=R_{\nu\mu}$ \cite{OSAT09,BO10}. In particular, the following action
\begin{equation}\label{eq:f(R,Q)}
S[g_{\mu\nu},\Gamma_{\alpha\beta}^\mu,\psi]=\frac{1}{2\kappa^2}\int d^4x \sqrt{-g}\left[R+a\frac{R^2}{R_P}+\frac{R_{\mu\nu}R^{\mu\nu}}{R_P}\right]+S_m[g_{\mu\nu},\psi] \ ,
\end{equation}
implies that $R=R(T)$ but $Q\equiv R_{\mu\nu}R^{\mu\nu}=Q(T_{\mu\nu})$, i.e., the scalar $Q$ has a more complicated dependence on the stress-energy tensor of matter than the trace. For instance, for a perfect fluid, one finds
\begin{equation}\label{eq:Q}
\frac{Q}{2R_P}=-\left(\kappa^2P+\frac{\tilde f}{2}+\frac{R_P}{8}\tilde f_R^2\right)+\frac{R_P}{32}\left[3\left(\frac{ R}{R_P}+\tilde f_R\right)-\sqrt{\left(\frac{R}{R_P}+\tilde f_R\right)^2-\frac{ 4 \kappa^2(\rho+P)}{R_P} }\right]^2 \ ,
\end{equation}
where $\tilde{f} =R+aR^2/R_P$ and $R$ is a solution of $R\tilde{f}_R-2\tilde{f}=\kappa^2 T$. From this it follows that
 even if one deals with a radiation fluid ($P=\rho/3$) or with a traceless field, the Palatini action (\ref{eq:f(R,Q)}) generates modified gravity without introducing new degrees of freedom. \\
For this model, it has been shown that completely regular bouncing solutions exist for both isotropic and anisotropic homogeneous cosmologies filled with a perfect fluid. In particular, one finds that for $a<0$ the interval $0\leq w\leq 1/3$ is always included in the family of bouncing solutions, which contains the dust and radiation cases. For $a\geq 0$, the fluids yielding a non-singular evolution are restricted to $w>\frac{a}{2+3a}$, which implies that the radiation case $w=1/3$ is always nonsingular. For a detailed discussion and classification of the non-singular solutions depending on the value of the parameter $a$ and the equation of state $w$, see \cite{BO10}. \\
As an illustration, consider a universe filled with radiation, for which $R=0$. In this case, the function $Q$ boils down to \cite{BO10}
\begin{equation}
Q= \frac{3R_P^2}{8}\left[1-\frac{8\kappa^2\rho}{3R_P}-\sqrt{1-\frac{16\kappa^2\rho}{3R_P}}\right] \label{eq:Q-rad} \ .
\end{equation}
This expression recovers the GR value at low curvatures, $Q\approx 4(\kappa^2\rho)^2/3+32(\kappa^2\rho)^3/9R_P+\ldots$ but reaches a maximum $Q_{max}=3R_P^2/16$ at $\kappa^2\rho_{max}=3R_P/16$, where the squared root of (\ref{eq:Q-rad}) vanishes. It can be shown that at $\rho_{max}$ the shear also takes its maximum, namely, $\sigma^2_{max}=\sqrt{3/16}R_P^{3/2}(C_{12}^2+C_{23}^2+C_{31}^2)$, which is always finite, and the expansion vanishes producing a cosmic bounce regardless of the amount of anisotropy (see Fig.\ref{fig:ExpanRad}). The model (\ref{eq:f(R,Q)}), therefore, avoids the well-known problems of anisotropic universes in GR, where anisotropies grow faster than the energy density during the contraction phase leading to a singularity that can only be avoided by sources with $w>1$.


\begin{thebibliography}{100}

\bibitem[Amelino-Camelia, 2002]{DSR1} Amelino-Camelia, G. (2002). \emph{ Int.J.Mod.Phys.} D{\bf 11}, 35.

\bibitem[Amelino-Camelia and Smolin, 2009]{AmelinoCamelia:2009pg}  Amelino-Camelia, G. and Smolin, L. (2009). \emph{Phys.\ Rev.\ } D {\bf 80}, 084017.

\bibitem[Ashtekar et al., 2006a]{APS06a} Ashtekar, A. , Pawlowski,T., and Singh,P. (2006). \emph{ Phys. Rev. Lett. } {\bf 96}, 141301.
\bibitem[Ashtekar et al., 2006b]{APS06b} Ashtekar, A. , Pawlowski,T., and Singh,P. (2006). \emph{ Phys. Rev.} {\bf D} 73,  124038.
\bibitem[Ashtekar et al., 2006c]{APS06c} Ashtekar, A. , Pawlowski,T., and Singh,P. (2006). \emph{ Phys. Rev. } {\bf D} 74, 084003.
\bibitem[Ashtekar, 2007]{Ash07} Ashtekar, A. (2007). \emph{ Nuovo Cim.} {\bf 122 B}, 135.

\bibitem[Barragan et al., 2009a]{BOSA09a} Barragan, C., Olmo, G.~J., and Sanchis-Alepuz, H. (2009). \emph{Phys.\ Rev.\ }  {\bf D80}, 024016,
  [arXiv:0907.0318 [gr-qc]].

\bibitem[Barragan et al., 2009b]{BOSA09b} Barragan, C., Olmo, G.~J., and Sanchis-Alepuz, H. (2009). [arXiv:1002.3919 [gr-qc]].

\bibitem[Barragan and Olmo, 2010]{BO10} Barragan, C., and Olmo, G.J. (2010). \emph{Phys. Rev. } {\bf D} 82, 084015.

\bibitem[Birkhoff, 1944]{Birkhoff} Birkhoff, G.D. (1944). Proc. Natl. Acad. Sci. U. S. A. 30(10), 324-334.

\bibitem[Bojowald, 2005]{Bojo05} Bojowald, M. (2005). \emph{ Living Rev. Rel.} {\bf 8}, 11.

\bibitem[Capozziello and Francaviglia, 2008]{Capo-Mauro} Capozziello, S. and Francaviglia, M. (2008). \emph{Gen. Rel. Grav.} 40, 357.

\bibitem[Copeland et al., 2006]{CSTs} Copeland, E.~J. et al. (2006). \emph{Int.\ J.\ Mod.\ Phys.\  {\bf D15}}, 1753-1936, [hep-th/0603057].

\bibitem[De Felice and Tsujikawa, 2010]{DFTs10}  De Felice, A. and Tsujikawa, S. (2010). \emph{Living Rev.\ Rel.\ }  {\bf 13}, 3, [arXiv:1002.4928 [gr-qc]].

\bibitem[Dicke, 1964]{Dicke1964} Dicke, R.H. (1964). \emph{The Theoretical Interpretation of Experimental Relativity}, Gordon and Breach, New York, U.S.A.

\bibitem[Einstein and Fokker, 1914]{Ein-Fok} Einstein, A. and Fokker, A.D. (1914). Ann. d. Phys. 44, 321.

\bibitem[Faraoni and Nadeau, 2005]{F-N05} Faraoni, V.  and Nadeau, S. (2005). \emph{Phys. Rev.} D 72, 124005.

\bibitem[Hawking, 1975]{Hawking-1975} Hawking, S.W. (1975). {\it Phys.Rev.} D {\bf 14}, 2460.

\bibitem[Magueijo, 2003]{VSL} Magueijo, J. (2003). {\it Rept. Prog. Phys.} {\bf 66}, 2025.

\bibitem[Magueijo and Smolin, 2002]{DSR2a} Magueijo, J. and Smolin, L. (2002). \emph{Phys.Rev.Lett.} {\bf 88}, 190403.

\bibitem[Magueijo and Smolin, 2003]{DSR2b} Magueijo, J. and Smolin, L. (2003). \emph{Phys.Rev.} D {\bf 67}, 044017.

\bibitem[Magueijo and Smolin, 2004]{Magueijo:2002xx} Magueijo, J. and Smolin, L. (2004). \emph{ Class. Quant. Grav.} {\bf 21}, 1725.


\bibitem[Norton, 1992]{Norton} Norton, J.D. (1992). Arch. Hist. Ex. Sci. 45, 17.

\bibitem[Nordstr\"{o}m, 1912]{Nord1} Nordstr\"{o}m, G. (1912). Phys. Zeit. 13, 1126.

\bibitem[Nordstr\"{o}m, 1913]{Nord2} Nordstr\"{o}m, G. (1913). Ann. d. Phys. 42, 533.

\bibitem[Novello and Bergliaffa, 2008]{Novello-2008} Novello, M. and Perez Bergliaffa, S.E. (2008). {\it Phys.Rep.} 463, 127-213.

\bibitem[Olmo, 2005a]{Olmo05a} Olmo, G.J. (2005). \emph{Phys. Rev.} D72, 083505.

\bibitem[Olmo, 2005b]{Olmo05b} Olmo, G.J. (2005). \emph{Phys. Rev. Lett.} 95, 261102.

\bibitem[Olmo, 2007]{Olmo07} Olmo, G.J. (2007). \emph{Phys. Rev.} D75, 023511.

\bibitem[Olmo and Singh, 2009]{OS-2009} Olmo, G.J. and Singh, P. (2009). JCAP 0901, 030.

\bibitem[Olmo et al., 2009]{OSAT09} Olmo, G.J., Sanchis-Alepuz, H. , and Tripathi, S. (2009). \emph{Phys. Rev. } {\bf D} 80, 024013.

\bibitem[Olmo, 2010]{Olmo10} Olmo,  G.J. (2010). AIP Conf.\ Proc.\  {\bf 1241}, 1100-1107, [arXiv:0910.3734 [gr-qc]].

\bibitem[Olmo, 2011a]{Olmo11a} Olmo, G.J. (2011). \emph{ Int. J. Mod. Phys. D}, 20, 413-462 (2011), [arXiv:1101.3864 [gr-qc]].

\bibitem[Olmo, 2011b]{Olmo11b}  Olmo, G.J. (2011).   JCAP\ {\bf 1110}, 018,   [arXiv:1101.2841 [gr-qc]].

\bibitem[Padmanabhan, 2003]{Pad03} Padmanabhan,  T. (2003). \emph{Phys. Rep.} 380, 235.

\bibitem[Peebles and Ratra, 2003]{PeRa03} Peebles, P. J. E. and Ratra, B. (2003). \emph{ Rev. Mod. Phys. }75, 559.

\bibitem[Rovelli, 2004]{Rovelli}  Rovelli, C. (2004).\emph{ Quantum Gravity}, Cambridge U. Press.

\bibitem[Singh, 2009]{Param09} Singh, P. (2009). \emph{ Class.\ Quant.\ Grav.\ } {\bf 26}, 125005, [arXiv:0901.2750 [gr-qc]].

\bibitem[Sotiriou and Faraoni, 2010]{SoFa08} Sotiriou, T.P.  and Faraoni, V. (2010). \emph{ Rev. Mod. Phys.} 82, 451-497, arXiv:0805.1726 [gr-qc].

\bibitem[Szulc et al., 2007]{SKL07} Szulc, L. , Kaminski, W. , and Lewandowski, J. (2007). \emph{ Class.Quant.Grav.} {\bf 24}, 2621.

\bibitem[Thiemann, 2007]{Thiemann} Thiemann, T. (2007).  \emph{ Modern canonical quantum general relativity}, Cambridge U. Press.

\bibitem[Will, 1993]{Will93} Will, C.M. (1993). \emph{Theory and Experiment in Gravitational Physics}, Cambridge University Press, Cambridge.

\bibitem[Will, 2005]{Will05} Will, C.M. (2005). {\it Living Rev.Rel.} 9,3,(2005), gr-qc/0510072 .

\bibitem[Zanelli, 2005]{Zanelli}  Zanelli, J. (2005). arXiv:hep-th/0502193v4 .


\end{thebibliography}
\end{document}